\documentclass[a4paper,12pt]{article}

\usepackage{amsmath}
\usepackage[psamsfonts]{amssymb}
\usepackage{rsfs}
\usepackage{bm}

\usepackage{cite}
\usepackage[dvips]{graphicx}
\usepackage{color}

\makeatletter
\@addtoreset{equation}{section}
\makeatother

\begin{document} 

\begin{titlepage}

\baselineskip 10pt
\hrule 
\vskip 5pt
\leftline{}
\leftline{Chiba Univ. Preprint
          \hfill   \small \hbox{\bf CHIBA-EP-176}}
\leftline{\hfill   \small \hbox{May 2009}}
\vskip 5pt
\baselineskip 14pt
\hrule 
\vskip 1.0cm
\centerline{\Large\bf 
} 
\vskip 0.3cm
\centerline{\Large\bf  
The nilpotent ``BRST'' symmetry
}
\vskip 0.3cm
\centerline{\Large\bf  
for the Gribov-Zwanziger theory
}
\vskip 0.3cm

\vskip 0.5cm

\centerline{{\bf 
Kei-Ichi Kondo,$^{\dagger,{1},{2}}$  
}}  
\vskip 0.5cm
\centerline{\it
${}^{1}$Department of Physics, University of Tokyo,
Tokyo 113-0033, Japan
}
\centerline{\it
${}^{2}$Department of Physics, 
Chiba University, Chiba 263-8522, Japan
}
\vskip 1cm

\begin{abstract}
We have found a nilpotent quantum symmetry of Yang-Mills theory restricted to the Gribov region. 
In fact, we give a set of transformations for the filds a la BRST that leaves the Gribov-Zwanziger action invariant and obeys the nilpotency, although the usual BRST symmetry is broken due to the non-vanishing Gribov parameter representing the presence of the Gribov horizon in the configuration space of the Yang-Mills field.
We show that the horizon condition (the gap equation determining the Gribov parameter) is consistent with the modified BRST invariance of the vacuum.
\end{abstract}

 
\vskip 0.5cm

\hrule  
\vskip 0.1cm
${}^\dagger$ 
On sabbatical leave of absence from Chiba University. 

  E-mail:  {\tt kondok@faculty.chiba-u.jp}

\par 
\par\noindent


\vskip 0.5cm

\pagenumbering{roman}
\tableofcontents




\end{titlepage}


\pagenumbering{arabic}

\baselineskip 14pt
\newpage
\section{Introduction and the main result}

The Kugo-Ojima criterion for color confinement \cite{KO79,Kugo95} is based on the assumption of an exact Becchi-Rouet-Stora-Tyupin (BRST) \cite{BRST} invariance of Yang-Mills theory in the manifestly covariant gauge.  In the Kugo-Ojima analysis, the color charge $Q^A$ of the global color current $\mathscr{J}^A$ plays the very important role in relation to the BRST symmetry generator, i.e., the BRST charge (operator) $Q_B$.  
In order to correctly quantize the Yang-Mills theory under the gauge fixing condition, however, one must avoid the Gribov copies in the Faddeev-Popov approach for Yang-Mills theory \cite{Gribov78}.  An available Lagrangian formalism for this purpose is the Gribov-Zwanziger theory \cite{Zwanziger89,Zwanziger92,Zwanziger93}.

 However, it is known that the restriction to the Gribov region for avoiding Gribov copies inevitably leads to a breaking of the BRST symmetry \cite{MS94,BS09,Dudaletal08}.
If the BRST symmetry is broken, the precise meaning of the Kugo-Ojima confinement criterion becomes unclear. 
Therefore, the Kugo-Ojima analysis cannot be applied to the Gribov-Zwanziger theory.  In view of this, it is not unreasonable that the Kugo-Ojima criterion can not be satisfied when the restriction to the Gribov region is taken into account, as shown numerically in \cite{FN07} and analytically in \cite{Kondo09}.  Consequently, the ghost propagator is no longer enhanced, in harmony with the latest lattice data \cite{lattice,Maas09} and other approaches \cite{Boucaudetal08,ABP08}.  

In spite of this fact, if we could find a modified BRST transformation which leaves the Gribov-Zwanziger action invariant, then we could apply the Kugo-Ojima analysis to the Gribov-Zwanziger theory, which opens the path to searching for the modified color confinement criterion {\it a la} Kugo and Ojima. 

In this paper, we obtain such a modified BRST symmetry of the Gribov-Zwanziger action.  The modified BRST transformation satisfies the nilpotency, but involves extra Gribov-parameter dependent terms, which preserve Lorentz covariance and color group structure, but lacks locality.

\section{Gribov-Zwanziger action}

It was shown \cite{Zwanziger89} that the partition function of the $D$-dimensional Euclidean Yang-Mills theory restricted to the first Gribov region for avoiding Gribov copies can be written in the form: 
\begin{equation}
 Z_{\gamma} := \int \mathcal{D}\mathscr{A} \delta (\partial^\mu \mathscr{A}_\mu) \det M \exp \{ -S_{YM} + \gamma \int d^D x h(x) \} 
 , 
 \label{YM1}
\end{equation}
where $S_{YM}$ is the Yang-Mills action, $M$ is the Faddeev-Popov operator $M:=-\partial_\mu D_\mu=-\partial_\mu (\partial_\mu+g \mathscr{A}_\mu \times)$ and $h(x)=h[\mathscr{A}](x)$ is the  horizon function given by  
\begin{equation}
 h(x) 
:= - \int d^Dy gf^{ABC} \mathscr{A}_\mu^{B}(x) (M^{-1})^{CE}(x,y) gf^{AFE} \mathscr{A}_\mu^{F}(y)
 . 
\end{equation}
Here the parameter $\gamma$ called the Gribov parameter is determined by solving a gap equation, commonly called the horizon condition:
\begin{equation}
 \langle h(x) \rangle^{\gamma} = (N^2-1)D .
\end{equation}
The action corresponding to the partition function (\ref{YM1}) contains the {\it non-local} horizon term: 
\begin{equation}
 \int d^Dx h(x) 
:= - \int d^Dx \int d^Dy gf^{ABC} \mathscr{A}_\mu^{B}(x) (M^{-1})^{CE}(x,y) gf^{AFE} \mathscr{A}_\mu^{F}(y)
 . 
\end{equation}

The Kugo-Ojima color confinement criterion was obtained in the framework of the BRST quantization for the usual Faddeev-Popov approach,  corresponding to the $\gamma=0$ case of the above Gribov-Zwanziger formulation.
In the conventional Faddeev-Popov approach, the $D$-dimensional Euclidean Yang-Mills theory in the covariant gauge with a gauge fixing parameter $\alpha$ is defined by
\begin{equation}
 Z  := \int [d\mathscr{A}] [d\mathscr{B}][d\mathscr{C}][d\bar {\mathscr{C}}]    \exp \{ -S_{YM}^{tot}  \} 
 , 
 \label{YM0}
\end{equation}
where
\begin{align}
 S_{YM}^{tot}  :=& S_{YM} + S_{GF+FP} ,
 \nonumber\\
 S_{YM} :=&   \int d^Dx \frac14 \mathscr{F}_{\mu\nu}  \cdot \mathscr{F}_{\mu\nu} ,
 \nonumber\\
 S_{GF+FP} :=& \int d^Dx  \left\{ \mathscr{B} \cdot \partial_\mu \mathscr{A}_\mu +  \frac{\alpha}{2} \mathscr{B} \cdot \mathscr{B} +i \bar {\mathscr{C}} \cdot \partial_\mu D_\mu \mathscr{C} \right\} ,
  \nonumber\\
 \mathscr{F}_{\mu\nu} :=& \partial_\mu \mathscr{A}_\nu - \partial_\nu \mathscr{A}_\mu +g \mathscr{A}_\mu \times \mathscr{A}_\nu ,
  \nonumber\\
 D_\mu \mathscr{C} :=& (\partial_\mu + g \mathscr{A}_\mu \times )  \mathscr{C}
 ,
\end{align}
and the dot and the cross are defined as 
\begin{equation}
  \mathscr{A} \cdot \mathscr{B} := \mathscr{A}^A \mathscr{B}^A , \quad 
  (\mathscr{A} \times \mathscr{B})^A := f^{ABC} \mathscr{A}^B \mathscr{B}^C 
  ,
\end{equation}
using the structure constant of the gauge group $G=SU(N)$.
The Landau gauge corresponds to $\alpha=0$. 
We assume that the Euclidean result is obtained by the Wick rotation of the Minkowski one.

Later, it has been shown \cite{Zwanziger92,Zwanziger93} that the non-local action (\ref{YM1}) can be put in an equivalent {\it local} form by introducing a set of complex conjugate commuting variables and anticommuting ones, which is called the Gribov-Zwanziger (GZ) action. 
The GZ theory is renormalizable to all orders of perturbation theory. Hence, the restriction to the (first) Gribov region $\Omega$ makes perfect sense at the quantum level, and finite results are obtained consistent with the renormalization group.

In order to fix the notation, we reproduce Gribov-Zwanziger (GZ) action.  
The non-local horizon term can be localized through the introduction of a suitable set of additional fields.

First, we prepare a set of commuting bosonic fields,  
$$
 \varphi_\mu^{AB}(x) \quad
(\mu=1, \cdots, D; \ 
A,B=1, \cdots, {\rm dim}G) .
$$
The Gaussian integration leads to
\begin{align}
& \int [d\varphi] \exp \left\{ 
 - \int d^Dx \left[
  \varphi_\mu^{CA} K^{AB} \varphi_\mu^{CB}  
+ 2i \gamma^{1/2} gf^{ABC} \mathscr{A}_\mu^B \varphi_\mu^{AC} 
 \right]
\right\} 
\nonumber\\
=& [(\det K)^{-1/2}]^{({\rm dim}G)D} 
\exp \left\{  
   \int d^D x  \int d^Dy \gamma (gf^{ABC} \mathscr{A}_\mu^B(x)) (K^{-1})^{CE}(x,y) (gf^{ADE} \mathscr{A}_\mu^D(y))
\right\} 
 ,
\end{align}
where we have defined the Faddeev-Popov (FP) operator by
\begin{equation}
 K^{AB} 
=:  - \partial_\mu D_\mu^{AB}[\mathscr{A}]
=: - \partial_\mu ( \partial_\mu \delta^{AB} + gf^{ACB} \mathscr{A}_\mu^{C} )
 .
\end{equation}
Hence, the horizon term is written as 
\begin{align}
& \exp \{ \gamma S_1[\mathscr{A}] \} 
\nonumber\\
=:& \exp \left\{  
   \int d^Dx  \int d^Dy \gamma (gf^{ABC} \mathscr{A}_\mu^B(x)) (K^{-1})^{CE}(x,y) (gf^{ADE} \mathscr{A}_\mu^D(y))
\right\} 
\nonumber\\
=& (\det K)^{({\rm dim}G)D/2} 
\int [d\varphi] \exp \left\{ 
 - \int d^Dx \left[
  \varphi_\mu^{CA} K^{AB} \varphi_\mu^{CB} 
+ 2i \gamma^{1/2} gf^{ABC} \mathscr{A}_\mu^B \varphi_\mu^{AC} 
 \right]
\right\} 
 .
\end{align}
Second, we incorporate the factor $(\det K)^{({\rm dim}G)D/2}$ into the action as an additional term. 
The procedure for realizing this goal is not unique. 
If we prepared a set of anticommuting independent Grassmannian field, $\omega_\mu^A, \bar{\omega}_\mu^A$ ($\mu=1, \cdots, D$;  $A=1, \cdots, {\rm dim}G$), then we would have
\begin{align}
&  (\det K)^{D} 
= 
\int [d\omega] [d\bar{\omega}] \exp \left\{ 
 - \int d^Dx \left[
  \bar{\omega}_\mu^{A} K^{AB} \omega_\mu^{B} 
 \right]
\right\} 
 ,
\end{align}
To achieve the correct factor $(\det K)^{({\rm dim}G)D/2}$, however, the anticommuting fields must have extra degrees of freedom represented by $\bar{\omega}_\mu^{CA}, \omega_\mu^{CB}$
($\mu=1, \cdots, D$;  $A,B,C=1, \cdots, {\rm dim}G$;  $C=1, \cdots, {\rm dim}G/2$):
\begin{align}
&  (\det K)^{({\rm dim}G)D/2} 
= 
\int [d\omega] [d\bar{\omega}] \exp \left\{ 
 - \int d^Dx \left[
  \bar{\omega}_\mu^{CA} K^{AB} \omega_\mu^{CB} 
 \right]
\right\} 
 ,
\end{align}
In this setting, however, there is an asymmetry between the two indices $A,B$ in $\bar{\omega}_\mu^{CA}, \omega_\mu^{CB}$. 
To  treat them on equal footing, therefore, we introduce $\bar{\omega}_\mu^{CA},\omega_\mu^{CB}$
($\mu=1, \cdots, D$;  $A,B,C=1, \cdots, {\rm dim}G$) to yield
\begin{align}
&  (\det K)^{({\rm dim}G)D} 
= 
\int [d\omega] [d\bar{\omega}] \exp \left\{ 
 - \int d^Dx \left[
  \bar{\omega}_\mu^{CA} K^{AB} \omega_\mu^{CB} 
 \right]
\right\} 
 ,
 \label{omega-integral}
\end{align}
where
$$
\omega_\mu^{CB}(x) , \bar{\omega}_\mu^{CA}(x)  \quad
(\mu=1, \cdots, D; \ 
A,B,C=1, \cdots, {\rm dim}G) .
$$
This treatment is preferable to consider the BRST transformation later.  There the mismatch of the indices between $\varphi$ and $\omega$ is not desirable.

In order to give rise to the extra factor $(\det K)^{({\rm dim}G)D/2}$, we need to introduce independent commuting field $\phi$.  Indeed, the horizon term can have another form:
\begin{align}
& \exp \{ \gamma S_1[\mathscr{A}] \} 
=  (\det K)^{({\rm dim}G)D} 
\nonumber\\
\nonumber\\
& \times
\int [d\varphi] [d\phi]  \exp \left\{ 
 - \int d^Dx \left[
  \varphi_\mu^{CA} K^{AB} \varphi_\mu^{CB} 
+ \phi_\mu^{CA} K^{AB} \phi_\mu^{CB} 
+ 2i \gamma^{1/2} gf^{ABC} \mathscr{A}_\mu^B \varphi_\mu^{AC} 
 \right]
\right\} 
 ,
\end{align}
Now we can use (\ref{omega-integral}) to rewrite the  horizon term in the local form:
\begin{equation}
 \exp \{ \gamma S_1[\mathscr{A}] \} 
=  
\int [d\varphi] [d\phi]  [d\omega] [d\bar{\omega}]  
 \exp \left\{   - S_\gamma[\mathscr{A},\varphi,\phi,\omega,\bar{\omega}]  \right\} 
 ,
\end{equation}
where
\begin{equation}
   S_\gamma  
=: \int d^Dx  \left[
  \varphi_\mu^{CA} K^{AB} \varphi_\mu^{CB} 
+ \phi_\mu^{CA} K^{AB} \phi_\mu^{CB} 
-  \bar{\omega}_\mu^{CA} K^{AB} \omega_\mu^{CB}
+ 2i \gamma^{1/2} gf^{ABC} \mathscr{A}_\mu^B \varphi_\mu^{AC} 
 \right] 
  .
 \label{S}  
\end{equation}

It is more convenient to introduce the complex field $\xi_\mu^{AB}$ and its complex conjugate $\bar{\xi}_\mu^{AB}$ by
\begin{equation}
 \xi_\mu^{AB} := \varphi_\mu^{AB} + i \phi_\mu^{AB} , \quad
 \bar{\xi}_\mu^{AB} := \varphi_\mu^{AB} - i \phi_\mu^{AB} 
  ,
\end{equation}
which implies the bilinear form (Note that the FP operator is Hermitian.)
\begin{align}
\bar{\xi}_\mu^{CA} K^{AB} \xi_\mu^{CB}
= \frac12 ( \bar{\xi}_\mu^{CA} K^{AB} \xi_\mu^{CB}+\xi_\mu^{CA} K^{AB} \bar{\xi}_\mu^{CB})   
= 
  \varphi_\mu^{CA} K^{AB} \varphi_\mu^{CB} 
+ \phi_\mu^{CA} K^{AB} \phi_\mu^{CB} 
  .
\end{align}
Hence, the horizon term has the local form:
\begin{equation}
 \exp \{ \gamma S_1[\mathscr{A}] \} 
=  
\int [d\xi] [d\bar{\xi}]  [d\omega] [d\bar{\omega}]  
 \exp \left\{   - \tilde{S}_\gamma[\mathscr{A},\xi,\bar{\xi},\omega,\bar{\omega}]  \right\} 
 ,
\end{equation}
where
\begin{equation}
    \tilde{S}_\gamma  
=: \int d^Dx  \left[
  \bar{\xi}_\mu^{CA} K^{AB} \xi_\mu^{CB} 
-  \bar{\omega}_\mu^{CA} K^{AB} \omega_\mu^{CB}
+ i \gamma^{1/2} gf^{ABC} \mathscr{A}_\mu^B  \xi_\mu^{AC} + i \gamma^{1/2} gf^{ABC} \mathscr{A}_\mu^B \bar{\xi}_\mu^{AC} 
 \right] 
  .  
  \label{aux-action}
\end{equation}
Thus, the localized action for the Gribov-Zwanziger theory reads
\begin{align}
  &  S_{\rm YM}^{\rm tot}[\mathscr{A},\mathscr{C},\bar{\mathscr{C}},\mathscr{B}] + \tilde{S}_\gamma [\mathscr{A},\xi,\bar{\xi},\omega,\bar{\omega}]  
  \nonumber\\
=& S_{\rm YM}[\mathscr{A}] + S_{\rm GF+FP}[\mathscr{A},\mathscr{C},\bar{\mathscr{C}},\mathscr{B}] 
+ \tilde{S}_\gamma [\mathscr{A},\xi,\bar{\xi},\omega,\bar{\omega}]
 ,
\end{align}
If $\gamma=0$, the factor resulting from the integration over $\xi$ and $\bar{\xi}$ cancel out that coming from the integration over  $\omega$ and $\bar{\omega}$.  Therefore, for $\gamma=0$, the introduction of $\tilde{S}_\gamma$ is trivial, which is equivalent to setting $\tilde{S}_\gamma=0$, and the usual BRST formulation of the Yang-Mills-Faddeev-Popov theory is reproduced.

\section{A modified BRST transformation}

Our main motivation is to find out a modified BRST transformation $\mbox{\boldmath $\delta$} ^\prime$ such that  $\mbox{\boldmath $\delta$} ^\prime$ leaves the action $S_{\rm YM}^{\rm tot}[\mathscr{A},\mathscr{C},\bar{\mathscr{C}},\mathscr{B}]+ \tilde{S}_\gamma [\mathscr{A},\xi,\bar{\xi},\omega,\bar{\omega}]$ invariant, i.e., 
\begin{equation}
     \mbox{\boldmath $\delta$} ^\prime (S_{\rm YM}^{\rm tot}[\mathscr{A},\mathscr{C},\bar{\mathscr{C}},\mathscr{B}] + \tilde{S}_\gamma [\mathscr{A},\xi,\bar{\xi},\omega,\bar{\omega}])   
= 0
 ,
\end{equation}
and  $\mbox{\boldmath $\delta$} ^\prime$ obeys the nilpotency, i.e., 
\begin{equation}
    \mbox{\boldmath $\delta$} ^\prime   \mbox{\boldmath $\delta$} ^\prime 
= 0
  .
\end{equation}
Such a transformation could be non-local.

For $\mathscr{A}_\mu^A(x)$, we require $\mbox{\boldmath $\delta$} ^\prime=\mbox{\boldmath $\delta$} $ to guarantee 
$
\mbox{\boldmath $\delta$}   (S_{\rm YM}[\mathscr{A}])
= 0
$, i.e.,
\begin{equation}
   \mbox{\boldmath $\delta$} ^\prime  \mathscr{A}_\mu
= \mbox{\boldmath $\delta$}   \mathscr{A}_\mu
= D_\mu[\mathscr{A}] \mathscr{C}
 .
\end{equation}
The requirement of the nilpotency
\begin{equation}
    \mbox{\boldmath $\delta$} ^\prime  \mbox{\boldmath $\delta$} ^\prime \mathscr{A}_\mu
= 0 
 ,
\end{equation}
uniquely determines the BRST transformation for the ghost field $\mathscr{C}$:
\begin{equation}
   \mbox{\boldmath $\delta$} ^\prime \mathscr{C}
= \mbox{\boldmath $\delta$}   \mathscr{C}
= - \frac{g}{2} (\mathscr{C} \times \mathscr{C})
 ,
\end{equation}
since 
\begin{equation}
    \mbox{\boldmath $\delta$} ^\prime  \mbox{\boldmath $\delta$} ^\prime \mathscr{A}_\mu
= D_\mu [\mbox{\boldmath $\delta$} ^\prime \mathscr{C}+\frac{g}{2} (\mathscr{C} \times \mathscr{C})] 
  .
\end{equation}
This transformation for the ghost field  automatically satisfies nilpotency:
\begin{equation}
   \mbox{\boldmath $\delta$} ^\prime  \mbox{\boldmath $\delta$} ^\prime \mathscr{C}
= 0 
  .
\end{equation}

For the antighost field $\bar{\mathscr{C}}$, we modify the BRST transformation as  
\begin{equation}
   \mbox{\boldmath $\delta$} ^\prime \bar{\mathscr{C}}
= \mbox{\boldmath $\delta$}  \bar{\mathscr{C}} + F
= i \mathscr{B} + F
 ,
\end{equation}
with $F$ being a functional to be determined later. 
Then, the nilpotency for the antighost field
\begin{equation}
    \mbox{\boldmath $\delta$} ^\prime  \mbox{\boldmath $\delta$} ^\prime \bar{\mathscr{C}}
= 0 
 ,
\end{equation}
determines the BRST transformation for the auxiliary field $\mathscr{B}$:
\begin{equation}
   \mbox{\boldmath $\delta$} ^\prime \mathscr{B}
= i \mbox{\boldmath $\delta$} ^\prime F
 ,
\end{equation}
which satisfies the nilpotency 
\begin{equation}
   \mbox{\boldmath $\delta$} ^\prime  \mbox{\boldmath $\delta$} ^\prime \mathscr{B}
= i \mbox{\boldmath $\delta$} ^\prime  \mbox{\boldmath $\delta$} ^\prime F
= 0
  .
\end{equation}

In order to determine the  functional $F$, we examine the GF+FP term of the form:
\begin{align}
 \mathscr{L}_{\rm GF+FP} :=  -i \mbox{\boldmath $\delta$} ^\prime \left[ \bar{\mathscr{C}}  \cdot  \left( \partial_\mu \mathscr{A}_\mu + \frac{\alpha}{2} \mathscr{B} \right) \right]  
  .
\end{align}
Applying the modified BRST transformation, we have
\begin{align}
 &  -i \mbox{\boldmath $\delta$} ^\prime \left[ \bar{\mathscr{C}} \cdot \left( \partial_\mu \mathscr{A}_\mu + \frac{\alpha}{2} \mathscr{B} \right) \right]  
 \nonumber\\
 =&   -i (\mbox{\boldmath $\delta$} ^\prime \bar{\mathscr{C}})  \cdot   \partial_\mu \mathscr{A}_\mu 
-i  \frac{\alpha}{2} (\mbox{\boldmath $\delta$} ^\prime  \bar{\mathscr{C}}) \cdot \mathscr{B}  
+ i \bar{\mathscr{C}}  \cdot \partial_\mu (\mbox{\boldmath $\delta$} ^\prime \mathscr{A}_\mu)
+ i \frac{\alpha}{2} \bar{\mathscr{C}}  \cdot  (\mbox{\boldmath $\delta$} ^\prime \mathscr{B})
\nonumber\\
=& 
  (  \mathscr{B} -i F)  \cdot   \partial_\mu \mathscr{A}_\mu 
+  \frac{\alpha}{2} ( \mathscr{B} -i F)  \cdot \mathscr{B}  
+ i \bar{\mathscr{C}}  \cdot \partial_\mu (D_\mu[\mathscr{A}] \mathscr{C})
+ i \frac{\alpha}{2}  \bar{\mathscr{C}} \cdot  (i \mbox{\boldmath $\delta$} ^\prime F)
 .
\end{align}
In the Landau gauge, $\alpha=0$, especially, the GF+FP term $S_{\rm GF+FP}[\mathscr{A},\mathscr{C},\bar{\mathscr{C}},\mathscr{B}]$ reduces to
\begin{align}
  \mathscr{L}_{\rm GF+FP} 
= -i \mbox{\boldmath $\delta$} ^\prime \left[ \bar{\mathscr{C}} \cdot  \partial_\mu \mathscr{A}_\mu   \right]  
=    \mathscr{B}  \cdot   \partial_\mu \mathscr{A}_\mu 
+ i \bar{\mathscr{C}}  \cdot \partial_\mu (D_\mu[\mathscr{A}] \mathscr{C})
+ i \partial_\mu  F \cdot   \mathscr{A}_\mu 
 ,
 \label{GFFPL}
\end{align}
where integration by parts is used in the last term. 

For the auxiliary fields $\xi$ and $\bar{\omega}$, we assume the modified BRST transformation of the form
\begin{equation}
   \mbox{\boldmath $\delta$} ^\prime \xi 
= \omega + G
 ,
 \label{BRST-G}
\end{equation}
\begin{equation}
   \mbox{\boldmath $\delta$} ^\prime \bar{\omega} 
= \bar{\xi} + H
  .
  \label{BRST-H}
\end{equation}
Then, the nilpotency for $\xi$ and $\bar{\omega}$, i.e., 
$
 \mbox{\boldmath $\delta$} ^\prime \mbox{\boldmath $\delta$} ^\prime \xi=0
$
and 
$
 \mbox{\boldmath $\delta$} ^\prime \mbox{\boldmath $\delta$} ^\prime \bar{\omega}  = 0
$
yield the BRST transformation for $\omega$ and $\bar{\xi}$:
\begin{equation}
   \mbox{\boldmath $\delta$} ^\prime \omega 
= - \mbox{\boldmath $\delta$} ^\prime G
 ,
\end{equation}
\begin{equation}
   \mbox{\boldmath $\delta$} ^\prime \bar{\xi} 
= - \mbox{\boldmath $\delta$} ^\prime H
  .
\end{equation}
This construction guarantees the nilpotency for   $\omega$ and $\bar{\xi}$, i.e., 
$
 \mbox{\boldmath $\delta$} ^\prime \mbox{\boldmath $\delta$} ^\prime \omega=0
$
and 
$
 \mbox{\boldmath $\delta$} ^\prime \mbox{\boldmath $\delta$} ^\prime \bar{\xi}  = 0
$.

In the Landau gauge, the functionals $F, G$ and $H$ are simultaneously determined by requiring that 
 additional part of the action 
$S_{\rm GF+FP}+ \tilde{S}_\gamma$
with 
$S_\gamma$ (\ref{aux-action}) is written in the BRST-exact form
\begin{align}
 & S_{\rm GF+FP}[\mathscr{A},\mathscr{C},\bar{\mathscr{C}},\mathscr{B}] 
+ \tilde{S}_\gamma [\mathscr{A},\xi,\bar{\xi},\omega,\bar{\omega}]
\nonumber\\
=& \int d^Dx   \{
   \mathscr{B}^A \partial_\mu \mathscr{A}_\mu^A 
- i \bar{\mathscr{C}}^A    K^{AB} \mathscr{C}^B 
+ \bar{\xi}_\mu^{CA} K^{AB} \xi_\mu^{CB} 
-  \bar{\omega}_\mu^{CA} K^{AB} \omega_\mu^{CB}
\nonumber\\
& \quad\quad\quad\quad + i \gamma^{1/2} gf^{ABC} \mathscr{A}_\mu^B  \xi_\mu^{AC} + i \gamma^{1/2} gf^{ABC} \mathscr{A}_\mu^B \bar{\xi}_\mu^{AC} 
 \}
\\
=& \int d^Dx   \{ -i \mbox{\boldmath $\delta$} ^\prime \left[ \bar{\mathscr{C}}^A (\partial_\mu \mathscr{A}_\mu^A)   \right] 
+ \mbox{\boldmath $\delta$} ^\prime  [\bar{\omega}_\mu^{CA} (-\partial_\rho D^{AB}_\rho[\mathscr{A}]) \xi_\mu^{CB}] \}
 ,
\end{align}
since the BRST invariance of $S_{\rm GF+FP} + \tilde{S}_\gamma$, i.e., $\mbox{\boldmath $\delta$} ^\prime(S_{\rm GF+FP} + \tilde{S}_\gamma)=0$, is guaranteed  by the nilpotency ($\mbox{\boldmath $\delta$} ^\prime \mbox{\boldmath $\delta$} ^\prime=0$) of the modified BRST transformation.
In view of this, we examine an exact form
\begin{align}
 \mathscr{L}_{\gamma} 
=& \mbox{\boldmath $\delta$} ^\prime  [\bar{\omega} K \xi]
\nonumber\\
=&  (\mbox{\boldmath $\delta$} ^\prime \bar{\omega}) K \xi 
- \bar{\omega} K (\mbox{\boldmath $\delta$} ^\prime \xi) 
- \bar{\omega} (\mbox{\boldmath $\delta$} ^\prime K) \xi 
\nonumber\\
=&  \bar{\xi}  K \xi 
+ H   K \xi 
- \bar{\omega} K \omega
- \bar{\omega} K G 
- \bar{\omega} (\mbox{\boldmath $\delta$} ^\prime K) \xi
\nonumber\\
=&  \bar{\xi}  K \xi 
- \bar{\omega} K \omega
+ H   K \xi 
- \bar{\omega}  [ KG +  (\mbox{\boldmath $\delta$} ^\prime K) \xi ]
  ,
\end{align}
where we have used (\ref{BRST-G}) and (\ref{BRST-H}). 
Therefore, we obtain the relationship:
\begin{subequations}
\begin{align}
KG  =& -  (\mbox{\boldmath $\delta$} ^\prime K) \xi ,
\label{pBRST-G}
\\
H   K \xi  =& i \gamma^{1/2} gf^{ABC} \mathscr{A}_\mu^B  \xi_\mu^{AC} ,
\label{pBRST-H}
\\
 \partial_\mu F^A \mathscr{A}_\mu^A =&    \gamma^{1/2} gf^{ABC} \mathscr{A}_\mu^B \bar{\xi}_\mu^{AC} 
  .  
\label{pBRST-F}
\end{align}
\end{subequations}
In this way, the modified BRST transformation for $\bar{\mathscr{C}}$, $\xi$ and $\bar{\omega}$ is determined.

A solution of the equation  (\ref{pBRST-F}) for $F(x)$, i.e.,
\begin{equation}
 \partial_\mu F^A   =  - \gamma^{1/2} gf^{ABC}  \bar{\xi}_\mu^{BC} 
  ,  
\end{equation}
is given by the line integral along an oriented path $L$ starting at somewhere, e.g., infinity for definiteness and ending at $x$:
\begin{equation}
   F^A(x)
=   - \gamma^{1/2}   gf^{ABC}  \int^{x}  dy_\mu \bar{\xi}_\mu^{BC}(y) 
  .
  \label{F-path}
\end{equation}
For this solution to be a path independent, namely, the solution $F(x)$ is determined uniquely at a spacetime point $x$ in question, the line integral along any closed path must be equal to zero, $\oint_C dy_\mu f^{ABC} \bar{\xi}_\mu^{BC}(y) =0$, which is equivalent to the rotation free (or integrability) condition of the vector field $v_\mu^A(x) := gf^{ABC} \bar{\xi}_\mu^{BC}(x)$, i.e.,
$\partial_\mu v_\nu^A(x) - \partial_\nu v_\mu^A(x) =0$.  Indeed, this is satisfied, since $v$ is of the exact one-form:
\begin{equation}
 v_\mu^A(x)  = gf^{ABC}  \bar{\xi}_\mu^{BC}(x)   = - \gamma^{-1/2}  \partial_\mu F^A(x)     
  .  
\end{equation}
However, we adopt another form for $F$: 
\begin{equation}
   F^A(x)
=  \gamma^{1/2} \int d^D y  \Delta^{-1}(x,y) \partial_\mu v_\mu^A(y)
=  \gamma^{1/2} \int d^D y  \Delta^{-1}(x,y)  gf^{ABC}  \partial_\mu \bar{\xi}_\mu^{BC}(y)  
 ,
\end{equation}
where $\Delta:=-\partial_\mu \partial_\mu$ is the $D$-dimensional Laplacian with the inverse
\begin{eqnarray}
\Delta^{-1}(x,y)
=   \frac{\Gamma\left(\frac{D}{2}-1 \right)}{4\pi^{D/2}|x - y|^{D-2}} .
\end{eqnarray}  
This form for $F^A$ is more suitable than the previous one (\ref{F-path}), since the path does not appear in the expression and it is similar to other functionals $G_\mu^{AB}$ and $H_\mu^{AB}$, which are obtained by operating the inverse of the FP operator from the left and right of (\ref{pBRST-G}) and (\ref{pBRST-H}), respectively. 

Thus, we have found the modified BRST transformation
\begin{align}
   \mbox{\boldmath $\delta$} ^\prime  \mathscr{A}_\mu^{A}(x) 
=&  (D_\mu[\mathscr{A}]\mathscr{C}(x))^{A}
 ,
 \nonumber\\
   \mbox{\boldmath $\delta$} ^\prime \mathscr{C}^{A}(x)
=&  -\frac{g}{2} (\mathscr{C}(x) \times \mathscr{C}(x))^{A}
 ,
 \nonumber\\
   \mbox{\boldmath $\delta$} ^\prime \bar{\mathscr{C}}^A(x)
=& i \mathscr{B}^A(x) + F^A(x)
 ,
 \nonumber\\
   \mbox{\boldmath $\delta$} ^\prime \mathscr{B}^{A}(x)
=&  i \mbox{\boldmath $\delta$} ^\prime F^{A}(x)
 ,
 \nonumber\\
   \mbox{\boldmath $\delta$} ^\prime \xi_\mu^{AB}(x) 
=& \omega_\mu^{AB}(x) + G_\mu^{AB}(x)
 ,
 \nonumber\\
    \mbox{\boldmath $\delta$} ^\prime \omega_\mu^{AB}(x) 
=&  - \mbox{\boldmath $\delta$} ^\prime G_\mu^{AB}(x)
 ,
 \nonumber\\
   \mbox{\boldmath $\delta$} ^\prime \bar{\omega}_\mu^{AB}(x) 
=& \bar{\xi}_\mu^{AB}(x) + H_\mu^{AB}(x)
 ,
 \nonumber\\
   \mbox{\boldmath $\delta$} ^\prime \bar{\xi}_\mu^{AB}(x) 
=&  - \mbox{\boldmath $\delta$} ^\prime H_\mu^{AB}(x)
 ,
\end{align}
where
\footnote{
The last equation can also be rewritten 
\begin{equation}
  H_\mu^{AB}(x) 
=   \int d^Dy i \gamma^{1/2} D_\mu^{BE}[\mathscr{A}](y)  (K^{-1})^{EA}(y,x)
  ,
\end{equation}
provided that the integral of the total derivative vanishes.
}
\begin{subequations}
\begin{align}
   F^A(x)
=&    \gamma^{1/2} \int d^D y  \Delta^{-1}(x,y)  gf^{ABC}  \partial_\mu \bar{\xi}_\mu^{BC}(y)   
 ,
\\
   G_\mu^{AB}(x) 
=& - \int d^Dy  (K^{-1})^{AC}(x,y) (\mbox{\boldmath $\delta$} ^\prime K)^{CE}(y) \xi_\mu^{EB}(y) 
 \nonumber\\
=&   \int d^Dy  (K^{-1})^{AC}(x,y) \partial_\rho [gf^{CFE} (D_\rho \mathscr{C})^{F}(y) \xi_\mu^{EB}(y) ]
 ,
\\
  H_\mu^{AB}(x) 
=&  \int d^Dy i \gamma^{1/2} gf^{BCE} \mathscr{A}_\mu^C(y)  (K^{-1})^{EA}(y,x)
  .
\end{align}
\end{subequations}
This modified BRST transformation is nilpotent and leaves the Gribov-Zwanziger action invariant. 
\begin{align}
 & S_{\rm GZ}[\mathscr{A},\mathscr{C},\bar{\mathscr{C}},\mathscr{B},\xi,\bar{\xi},\omega,\bar{\omega}] 
\nonumber\\
=& S_{\rm YM}[\mathscr{A}] + S_{\rm GF+FP}[\mathscr{A},\mathscr{C},\bar{\mathscr{C}},\mathscr{B}] 
+ \tilde{S}_\gamma [\mathscr{A},\xi,\bar{\xi},\omega,\bar{\omega}]
\nonumber\\
=& S_{\rm YM}[\mathscr{A}] + \int d^Dx   \{ -i \mbox{\boldmath $\delta$} ^\prime \left[ \bar{\mathscr{C}}^A (\partial_\mu \mathscr{A}_\mu^A)   \right] 
+ \mbox{\boldmath $\delta$} ^\prime  [\bar{\omega}_\mu^{CA} (-\partial_\rho D^{AB}_\rho[\mathscr{A}]) \xi_\mu^{CB}] \}
  .
\end{align}
Among the fields, $\mathscr{A},\mathscr{C},\bar{\mathscr{C}},\mathscr{B},\xi,\bar{\xi},\omega,\bar{\omega}$, only the fields $\mathscr{A},\bar{\mathscr{C}},\xi,\bar{\omega}$ appear in this construction.  Hence, the BRST transformation for the fields $\mathscr{C},\mathscr{B},\bar{\xi},\omega$ do not affect the action. They are determined by the nilpotency requirement. 

In the limit $\gamma \to 0$,   
$F^A(x) \to 0$, $H_\mu^{AB}(x) \to 0$
\begin{align}
   \mbox{\boldmath $\delta$} ^\prime  \mathscr{A}_\mu^{A}(x) 
=&  (D_\mu[\mathscr{A}]\mathscr{C}(x))^{A}
 ,
 \nonumber\\
   \mbox{\boldmath $\delta$} ^\prime \mathscr{C}^{A}(x)
=&  -\frac{g}{2} (\mathscr{C}(x) \times \mathscr{C}(x))^{A}
 ,
 \nonumber\\
   \mbox{\boldmath $\delta$} ^\prime \bar{\mathscr{C}}^A(x)
=& i \mathscr{B}^A(x)  
 ,
 \nonumber\\
   \mbox{\boldmath $\delta$} ^\prime \mathscr{B}^{A}(x)
=& 0
 ,
 \nonumber\\
   \mbox{\boldmath $\delta$} ^\prime \xi_\mu^{AB}(x) 
=& \omega_\mu^{AB}(x) + G_\mu^{AB}(x)
 ,
 \nonumber\\
    \mbox{\boldmath $\delta$} ^\prime \omega_\mu^{AB}(x) 
=&  - \mbox{\boldmath $\delta$} ^\prime G_\mu^{AB}(x)
 ,
 \nonumber\\
   \mbox{\boldmath $\delta$} ^\prime \bar{\omega}_\mu^{AB}(x) 
=& \bar{\xi}_\mu^{AB}(x) 
 ,
 \nonumber\\
   \mbox{\boldmath $\delta$} ^\prime \bar{\xi}_\mu^{AB}(x) 
=&  0
 .
\end{align}

Note that $G$ does not vanish even in the  limit $\gamma \to 0$ and the modified BRST transformation $\mbox{\boldmath $\delta$} ^\prime \xi_\mu^{AB}(x)$ has the part $G$ involving the Yang-Mills field and the ghost field. 
Even in the limit, therefore, the horizon term is not decoupled from the usual Yang-Mills-Faddeev-Popov theory.
This issue is cured by redefining the auxiliary field $\omega_\mu^{AB}(x)$, i.e., shifting it by $G_\mu^{AB}(x)$  \cite{Zwanziger93}:
\begin{equation}
 \omega^\prime{}_\mu^{AB}(x) := \omega_\mu^{AB}(x) + G_\mu^{AB}(x)
  .
\end{equation}
Then the modified BRST transformation is simplified 
\begin{subequations}
\begin{align}
   \mbox{\boldmath $\delta$} ^\prime  \mathscr{A}_\mu^{A}(x) 
=&  (D_\mu[\mathscr{A}]\mathscr{C}(x))^{A}
 ,
\\
   \mbox{\boldmath $\delta$} ^\prime \mathscr{C}^{A}(x)
=&  -\frac{g}{2} (\mathscr{C}(x) \times \mathscr{C}(x))^{A}
 ,
\\
   \mbox{\boldmath $\delta$} ^\prime \bar{\mathscr{C}}^A(x)
=& i \mathscr{B}^A(x) + F^A(x)
 ,
\\
   \mbox{\boldmath $\delta$} ^\prime \mathscr{B}^{A}(x)
=&  i \mbox{\boldmath $\delta$} ^\prime F^{A}(x)
 ,
\\
   \mbox{\boldmath $\delta$} ^\prime \xi_\mu^{AB}(x) 
=& \omega^\prime{}_\mu^{AB}(x)  
 ,
\\
    \mbox{\boldmath $\delta$} ^\prime \omega^\prime{}_\mu^{AB}(x) 
=&  0
 ,
\\
   \mbox{\boldmath $\delta$} ^\prime \bar{\omega}_\mu^{AB}(x) 
=& \bar{\xi}_\mu^{AB}(x) + H_\mu^{AB}(x)
 ,
\\
   \mbox{\boldmath $\delta$} ^\prime \bar{\xi}_\mu^{AB}(x) 
=&  - \mbox{\boldmath $\delta$} ^\prime H_\mu^{AB}(x)
 ,
\end{align}
where
\begin{align}
   F^A(x)
=&   \gamma^{1/2} \int d^D y  \Delta^{-1}(x,y)  gf^{ABC}  \partial_\mu \bar{\xi}_\mu^{BC}(y)  
 ,
\\
  H_\mu^{AB}(x) 
=&  \int d^Dy i \gamma^{1/2} gf^{CEB} \mathscr{A}_\mu^C(y)  (K^{-1})^{EA}(y,x)
  .
\end{align}
\end{subequations}
Here note that the BRST transformation is closed, although it is quite non-local, as can be seen as follows.
\begin{align}
  \mbox{\boldmath $\delta$} ^\prime F^A(x)
=&  \gamma^{1/2} \int d^D y  \Delta^{-1}(x,y)  gf^{ABC}  \partial_\mu \mbox{\boldmath $\delta$} ^\prime \bar{\xi}_\mu^{BC}(y) 
 ,
\\
  \mbox{\boldmath $\delta$} ^\prime  H_\mu^{AB}(y) 
=&  \int d^Dz i \gamma^{1/2} gf^{CEB} [\mbox{\boldmath $\delta$} ^\prime  \mathscr{A}_\mu^C(z)  (K^{-1})^{EA}(z,y)
+  \mathscr{A}_\mu^C(z)  \mbox{\boldmath $\delta$} ^\prime   (K^{-1})^{EA}(z,y) ]
  .
\end{align}
It turns out that the nilpotency is satisfied for all the fields.  In fact, the nilpotency reduces to that for $\mathscr{A}_\mu^A$:
\begin{align}
  \mbox{\boldmath $\delta$} ^\prime \mbox{\boldmath $\delta$} ^\prime \mathscr{B}^{A}(x) 
=&   i \mbox{\boldmath $\delta$}^\prime  \mbox{\boldmath $\delta$}^\prime F^A(x)
 \nonumber\\
=& i \gamma^{1/2} \int d^D y  \Delta^{-1}(x,y)  gf^{ABC}  \partial_\mu  \mbox{\boldmath $\delta$}^\prime \mbox{\boldmath $\delta$}^\prime  \bar{\xi}_\mu^{BC}(y) 
 ,
\\
 \mbox{\boldmath $\delta$} ^\prime \mbox{\boldmath $\delta$} ^\prime \bar{\xi}_\mu^{AB}(x) 
=&  -\mbox{\boldmath $\delta$} ^\prime \mbox{\boldmath $\delta$} ^\prime  H_\mu^{AB}(y) 
 \nonumber\\
=& - \int d^Dz i \gamma^{1/2} gf^{CEB} [ \mathscr{A}_\mu^C(z)   \mbox{\boldmath $\delta$} ^\prime \mbox{\boldmath $\delta$} ^\prime   (K^{-1})^{EA}(z,y) ]
= 0
  .
\end{align}
In the limit $\gamma \to 0$, the horizon term is decoupled from the usual Yang-Mills-Faddeev-Popov theory:
\begin{align}
   \mbox{\boldmath $\delta$} ^\prime  \mathscr{A}_\mu^{A}(x) 
=&  (D_\mu[\mathscr{A}]\mathscr{C}(x))^{A}
 ,
 \nonumber\\
   \mbox{\boldmath $\delta$} ^\prime \mathscr{C}^{A}(x)
=&  -\frac{g}{2} (\mathscr{C}(x) \times \mathscr{C}(x))^{A}
 ,
 \nonumber\\
   \mbox{\boldmath $\delta$} ^\prime \bar{\mathscr{C}}^A(x)
=& i \mathscr{B}^A(x)  
 ,
 \nonumber\\
   \mbox{\boldmath $\delta$} ^\prime \mathscr{B}^{A}(x)
=&  0
 ,
 \nonumber\\
   \mbox{\boldmath $\delta$} ^\prime \xi_\mu^{AB}(x) 
=& \omega^\prime{}_\mu^{AB}(x)  
 ,
 \nonumber\\
    \mbox{\boldmath $\delta$} ^\prime \omega^\prime{}_\mu^{AB}(x) 
=&  0
 ,
 \nonumber\\
   \mbox{\boldmath $\delta$} ^\prime \bar{\omega}_\mu^{AB}(x) 
=& \bar{\xi}_\mu^{AB}(x)  
 ,
 \nonumber\\
   \mbox{\boldmath $\delta$} ^\prime \bar{\xi}_\mu^{AB}(x) 
=&   0
 .
\end{align}
For this choice of the field variables, the action changes  \cite{Zwanziger93}
\begin{align}
 & S_{\rm GF+FP}[\mathscr{A},\mathscr{C},\bar{\mathscr{C}},\mathscr{B}] 
+ \tilde{S}_\gamma [\mathscr{A},\xi,\bar{\xi},\omega^\prime,\bar{\omega}]
\nonumber\\
=& \int d^Dx   \{ -i \mbox{\boldmath $\delta$} ^\prime \left[ \bar{\mathscr{C}}^A (\partial_\mu \mathscr{A}_\mu^A)   \right] 
+ \mbox{\boldmath $\delta$} ^\prime  [\bar{\omega}_\mu^{CA} (-\partial_\rho D^{AB}_\rho[\mathscr{A}]) \xi_\mu^{CB}] \}
\nonumber\\
=& \int d^Dx   \{
\mathscr{B}^A   \partial_\mu \mathscr{A}_\mu^A 
+  i \bar{\mathscr{C}}^A \partial_\mu (D_\mu^{AB}[\mathscr{A}] \mathscr{C}^B)
\nonumber\\
&\quad\quad - \bar{\xi}_\mu^{CA} \partial_\rho D^{AB}_\rho[\mathscr{A}] \xi_\mu^{CB} 
+  \bar{\omega}_\mu^{CA} \partial_\rho D^{AB}_\rho[\mathscr{A}] \omega^\prime{}_\mu^{CB}
-  \bar{\omega}_\mu^{CA} \partial_\rho [gf^{CDE} (D_\rho \mathscr{C})^{D}   \xi_\mu^{EA}]
\nonumber\\
& \quad\quad
 + i \gamma^{1/2} gf^{ABC} \mathscr{A}_\mu^B  \xi_\mu^{AC} + i \gamma^{1/2} gf^{ABC} \mathscr{A}_\mu^B \bar{\xi}_\mu^{AC} 
 \}
 .
\end{align}
The Jacobian for the non-local change of variable from $\omega_\mu^{AB}(x)$ to $\omega^\prime{}_\mu^{AB}(x)$ is field-independent and hence the change of variable does not generate additional factor in the integration measure or the action.

It is possible to further simplify the BRST transformation by the redefinition of the field $\bar{\xi}^\prime=\bar{\xi}+H$, 
$\mathscr{B}^\prime=\mathscr{B}-iF$.
On the other hand, the integration measure may be complicated due to the Jacobian associated with the non-local change of variables. 
Consequently, such an option of the Gribov-Zwanziger theory would contain non-local terms again. 

In order to demonstrate the usefulness of the modified BRST symmetry, we show that the horizon condition  (the gap equation determining the Gribov parameter) is consistent with the BRST invariance of the vacuum. 
We define the generator $Q_B^\prime$, i.e., the BRST charge of the modified BRST transformation by
\begin{equation}
 \mbox{\boldmath $\delta$} ^\prime \Phi(x) = [iQ_B^\prime, \Phi(x)]_{\mp} 
\end{equation} 
where $ [, ]_{\mp}$ denotes the commutator or anticommutator depending on the ghost number of $\Phi$ being even or odd, respectively. 
Then we have an Slavnov-Taylor identity:
\begin{align}
 & \langle 0| [iQ_B^\prime, gf^{ABC}\mathscr{A}_\mu^A(x) \bar{\omega}^{BC}_\mu(x)]_{+}  | 0 \rangle 
\nonumber\\
 =& \langle 0| \mbox{\boldmath $\delta$} ^\prime [gf^{ABC}\mathscr{A}_\mu^A(x) \bar{\omega}^{BC}_\mu(x)]  | 0 \rangle 
\nonumber\\
 =&  \langle 0| [gf^{ABC} \mbox{\boldmath $\delta$} ^\prime \mathscr{A}_\mu^A(x) \bar{\omega}^{BC}_\mu(x)]  | 0 \rangle
 + \langle 0|  [gf^{ABC}\mathscr{A}_\mu^A(x) \mbox{\boldmath $\delta$} ^\prime  \bar{\omega}^{BC}_\mu(x)]  | 0 \rangle
\nonumber\\
 =&  \langle 0| [gf^{ABC} (D_\mu \mathscr{C})^A(x) \bar{\omega}^{BC}_\mu(x)]  | 0 \rangle
 + \langle 0|  [gf^{ABC}\mathscr{A}_\mu^A(x) \bar{\xi}^{BC}_\mu(x)]  | 0 \rangle
\nonumber\\
& + \langle 0|  [gf^{CAB}\mathscr{A}_\mu^A(x) \int d^D y i \gamma^{1/2} gf^{CEF} \mathscr{A}_\mu^{E}(y) (K^{-1})^{FB}(y,x)]  | 0 \rangle
\nonumber\\
 =&   \langle 0|  [gf^{ABC}\mathscr{A}_\mu^A(x) \bar{\xi}^{BC}_\mu(x)]  | 0 \rangle
+ i \gamma^{1/2} \langle 0|  h(x) | 0 \rangle
  ,
\end{align}
where we have used the fact that 
$
\langle 0| [gf^{ABC} (D_\mu \mathscr{C})^A(x) \bar{\omega}^{BC}_\mu(x)]  | 0 \rangle
$
vanishes due to the discrete symmetry of the Gribov-Zwanziger action, 
\begin{equation}
 \bar{\omega}_\mu \rightarrow - \bar{\omega}_\mu, \quad 
 \omega_\mu \rightarrow - \omega_\mu 
  .
\end{equation} 
The original horizon condition in the non-local Zwanziger theory \cite{Zwanziger89} is given by
$
 \langle 0|  h(x) | 0 \rangle = (N^2-1)D
$.
Therefore, the horizon condition in the Gribov-Zwanziger theory \cite{Zwanziger93}
\begin{equation}
 \langle 0|  [gf^{ABC}\mathscr{A}_\mu^A(x) \bar{\xi}^{BC}_\mu(x)]  | 0 \rangle
 = -i\gamma^{1/2} (N^2-1)D  
  ,
\end{equation} 
follows from the BRST invariance of the vacuum:
\begin{equation}
 Q_B^\prime  | 0 \rangle 
= 0 ,
\end{equation} 
which yields
\begin{equation}
\langle 0| [iQ_B^\prime, gf^{ABC}\mathscr{A}_\mu^A(x) \bar{\omega}^{BC}_\mu(x)]_{+}  | 0 \rangle 
= 0 .
\end{equation} 
In other words, if the vacuum in the Gribov-Zwanziger theory obeys the subsidiary condition
$Q_B^\prime  | 0 \rangle = 0$, the horizon condition is automatically satisfied.

Finally, it should be remarked that the Gribov-Zwanziger theory can be applied only to the Landau gauge $\alpha=0$ in the manifestly covariant gauge, which restricts the gauge field to the hypersurface $\partial_\mu \mathscr{A}_\mu^A(x)=0$, while  for $\alpha \ne 0$ the gauge field is allowed to be anywhere in the configuration space of the field.
The explicit form of the modified BRST transformation obtained in this paper is meaningful only in the Landau gauge. If one considers the other gauge fixing, e.g., the Coulomb gauge or the Maximal Abelian gauge in which the Gribov copies are known to exist, the corresponding Gribov-Zwanziger theory has the different action depending on the gauge fixing, and hence the modified ``BRST'' transformation obtained in the same sense as that in this paper may differ gauge to gauge.  On the contrary, the usual BRST transformation is independent of the gauge fixing condition to be chosen.  In this sense, the modified BRST transformation in this paper should be understood as a quantum symmetry for the respective theory,  which is non-local and nilpotent. 
The modified confinement criterion remains to be clarified, once the nilpotent BRST symmetry has been found. 

[Note added]
In preparing this paper, we have found that a paper by Sorella \cite{Sorella09} was submitted to the archive.  In this paper, he obtained a modified BRST transformation leaving the Gribov-Zwanziger action invariant.  As expected, the transformation is non-local in the sense that the inverse of the FP operator is involved in it.  However, the modified BRST transformations lack nilpotency.  Our BRST transformation satisfies nilpotency by construction. 
The usefulness of the non-local BRST transformation is demonstrated in  his paper.

\section*{Acknowledgments}
The author would like to thank High Energy Physics Theory Group and Theoretical Hadron Physics Group in the University of Tokyo, especially, Prof. Tetsuo Hatsuda for kind hospitality extended to him on sabbatical leave.
He would like to thank Dr. David Dudal and Dr. Nele Vandersickel for sending valuable comments and questions on the first version of this paper. 
This work is financially supported by Grant-in-Aid for Scientific Research (C) 21540256 from Japan Society for the Promotion of Science
(JSPS).


\end{document}